\documentclass[aps,pra,twocolumn,showpacs]{revtex4}
\usepackage{graphicx}
 
\newcommand{\boldvec}[1]{\mbox{\boldmath$#1$}}
\newcommand{\smallvec}[1]{\mbox{\boldmath$\scriptstyle#1$}}
\newcommand{\fdestroy}[2]{a^{\phantom *}_{#1}({#2})}
\newcommand{\fcreate}[2]{a^{*}_{#1}({#2})}
\newcommand{\bdestroy}[1]{b_j({#1})}
\newcommand{\bcreate}[1]{b_j^*({#1})}

\begin{document}
 

\title{Resonance theory of the crossover from
  Bardeen-Cooper-Schrieffer superfluidity to Bose-Einstein
  condensation in a dilute Fermi gas}
 
\author{J. N. Milstein, S. J. J. M. F. Kokkelmans, and M. J. Holland}
 
\affiliation{JILA, University of Colorado and National Institute of
  Standards and Technology, Boulder, Colorado 80309-0440}

\begin{abstract}
  We present a description of the behavior of a superfluid gas of
  fermions in the presence of a Feshbach resonance over the complete
  range of magnetic field detunings.  Starting from a resonance
  Hamiltonian, we exploit a functional method to describe the
  continuous behavior from Bardeen-Cooper-Schrieffer to
  Bose-Einstein condensation type superfluidity. Our results
  show an ability for a resonance system to exhibit a high critical
  temperature comparable to the Fermi temperature. The results are
  derived in a manner that is shown to be consistent with the
  underlying microscopic scattering physics.
\end{abstract}
 
\pacs{03.75.Fi,64.60.-i,74.20.-z}
\maketitle

\section{Introduction}

The ability to cool a gas of fermionic atoms well into the regime of quantum
degeneracy hints at the exciting possibility of allowing one to
study the mechanisms of superfluidity in an entirely new context \cite{demarco}.
Since these systems remain extremely dilute and are not complicated by
long-range Coulomb interactions or lattice effects, cold degenerate
gases seem ideal for the study of the fundamental physics behind the
exotic behavior of superfluidity. Unfortunately, temperatures of around
$0.2T_F$ are the current state of the art in cooling \cite{demarco,hadzibabic,truscott,schreck,granade,inguscio}. To obtain such high critical temperatures a strong coupling mechanism is required forcing the theoretical description to extend beyond the standard Bardeen-Cooper-Schrieffer (BCS) approach. 

Several theoretical papers have studied the
effects of increasing the two-particle interactions~\cite{stoof,combescot,heiselberg}, characterizing the
coupling processes by large negative scattering lengths. We have
focused in detail upon a related yet distinct approach
\cite{servaas}, which is to significantly increase the interatomic
couplings by making use of a Feshbach resonance.
The difference is that, in the neighborhood of the resonance, the
interactions can no longer be adequately described by a scattering
length, since the scattering length diverges as one approaches the
resonance. This is an artifact of the approximations made in
formulating the theory since the full energy-dependent scattering T-matrix, which is
the true descriptor of the two-particle interactions, does not diverge at finite scattering energy.  This leads us to explicitly incorporate the
physics of the resonance into our microscopic description of the
interatomic couplings. A much more detailed discussion of this can be
found in Ref.~\cite{servaas}.

So far, we have not discussed in detail the role of
fluctuations, which can have a significant effect on the critical behavior~\cite{gorkov}. How we incorporate these fluctuations proves crucial in describing the
physics correctly within the crossover regime where we find a significant
population of tightly bound composite particles. The aim, therefore,
of this paper will be to account for fluctuations in such a way as to
properly describe the behavior of a superfluid Fermi gas at all
detunings from the resonance. Recently, a complimentary treatment was independently developed by Ohashi and Griffin \cite{ohashi}.  The slight quantitative differences between their results for the critical superfluid transition temperature and the values we will present here appear to arise primarily from the use of quite different two-body scattering parameters and a distinct renormalization procedure.

The problem of describing a superfluid Fermi gas at all coupling strengths
 has been extensively studied in recent years, motivated
by a desire to explain the properties of ``exotic'' high-$T_c$
superconductors, whose behavior seems to lie in a region somewhere
between BCS and Bose-Einstein condensation (BEC) superconductivity. An early description of the
crossover from BCS to BEC superconductivity was put forth by
Nozi\`eres and Schmitt-Rink (NSR) \cite{nozieres}, after the
pioneering work of Eagles \cite{eagles} and Leggett \cite{leggett},
and later expanded upon by various authors
\cite{miscross,haussmann,randeria}.  A functional analysis of the
crossover behavior, which is the method that we will employ, was
equated to the NSR method by Randeria {\it et al.}~\cite{randeria}.
  We will adapt this method to a resonant
system as necessary to describe the relevant physics of superfluidity
in dilute atomic gases. It should be stressed that our method contains
the multichannel interatomic couplings intrinsic
to the Feshbach resonance. This was not considered
in previous calculations in the context of condensed matter systems.

\section{Resonant Action}

We consider the Feshbach resonance \cite{verhaar} for $s$-wave scattering of atoms in
the lowest two hyperfine states of a fermionic alkali atom, denoted symbolically by
\mbox{$\sigma\in\{\uparrow,\downarrow\}$}.  For a homogeneous system we have
the following generalized Hamiltonian:
\begin{eqnarray}
\hat{H}(t)&=&\sum_\sigma\int\psi_\sigma^\dagger(x)
(\hat{H}_\sigma-\mu) \psi_\sigma(x) d^3\boldvec{x}\\
&&\hspace{-10mm}+\sum_j\int\psi_{m_j}^\dagger(x)
(\hat{H}_{m_j}-2\mu+\nu_j) \psi_{m_j}(x) d^3\boldvec{x}\nonumber\\
&&\hspace{-10mm}+\int U(\boldvec{x}-\boldvec{x'})
\psi_\uparrow^\dagger(x)\psi_\downarrow^\dagger(x')\psi_\downarrow(x')\psi_\uparrow(x) d^3\boldvec{x}d^3\boldvec{x'}\nonumber\\
&&\hspace{-20mm}+\sum_j\int\left(g_j(\boldvec{x}-\boldvec{x'})\psi_{m_j}^\dagger(\frac{x+x'}{2})
\psi_\downarrow(x)\psi_\uparrow(x')+{\rm H.c.} \right)d^3
\boldvec{x}d^3\boldvec{x'},\label{ham}\nonumber
\end{eqnarray}
where the operators $\psi_\sigma^\dagger$ ($\psi_\sigma$)
create (annihilate) fermions at \mbox{$x=(\boldvec{x},t)$}, and
$\psi_{m_j}^\dagger$ ($\psi_{m_j}$) create (annihilate) composite
bosons. The free dispersion Hamiltonian for fermions (bosons) is
$\hat{H}_\sigma$ ($\hat{H}_{m_j}$) and $\nu_j$ is the detuning of the
$j^{\rm th}$ molecular state from the collision continuum.  The
collisional interactions are described by both background fermion
scattering ($U$) and an interconversion between composite bosons and
fermion pairs ($g_j$).

Functional methods prove to be especially convenient in describing the
thermodynamics of the resonant system.  For a finite temperature field
theory, the connection with statistical mechanics is made by Wick
rotating the time coordinate $t\rightarrow -i\tau$ so
that one works in terms of the spatial coordinate $x$ and temperature
$\tau$ \cite{abrikosov}.  In this space, we define the action in the usual way
\begin{eqnarray}
S=\sum_l\int_0^\beta d\tau\int d^3\boldvec{x}
\psi_l^\dagger(\boldvec{x},\tau)\partial_\tau\psi_l
(\boldvec{x},\tau)-\int_0^\beta \hat{H}(\tau)d\tau,\nonumber\\
\end{eqnarray}
where the sum in $l$ runs over both the Fermi and the Bose degrees of
freedom. In this functional formulation we treat
the fermion fields $\psi_\sigma$ as Grassmann variables \cite{popov}
and the composite Bose fields $\psi_{m_j}$ as classical fields.

Let us consider a system comprised of fermions at some finite
temperature $\tau$ inside a box of volume V (for convenience, let us
work in the set of units where $\hbar=k_b=1$).  By imposing periodic
boundary conditions upon the fields $\psi_\sigma$ and $\psi_{m_j}$, we
form the following Fourier series expansions
\begin{eqnarray}
\psi_\sigma(\boldvec{x},\tau)&=&(\beta V)^{-1/2}
\sum_{\smallvec{k},\omega}e^{i(\omega\tau+\smallvec{p}\cdot\smallvec{x})}
\fdestroy{\sigma}{\rm{p}}\nonumber,\\
\psi_{m_j}(\boldvec{x},\tau)&=&(\beta V)^{-1/2}
\sum_{\smallvec{q},v}e^{i(v\tau+\smallvec{q}\cdot\smallvec{x})}
\bdestroy{\rm{q}}\label{four},
\end{eqnarray}
with even thermal (Matsubara) frequencies for the bosons ($v=2\pi
n/\beta$, where $n$ is an integer) and odd frequencies for the
fermions ($\omega=2\pi (n+1)/\beta$), to preserve the particle
statistics. Here $\fdestroy{\sigma}{\rm{p}}$ annihilates a fermion at $\rm{p}=(\boldvec{k},\omega)$ and $\bdestroy{\rm{q}}$ annihilates a molecule at $\rm{q}=(\boldvec{q},\omega)$.

By making use of the above transformation, Eq.~(\ref{four}), we may
write out the action for the resonant system in terms of the Fourier
coefficients $a_\sigma(\rm{p})$ and $b_j(\rm{q})$. In order to help clarify the
following calculation, we split the resulting resonant action into two
parts, the first being the usual BCS action:
\begin{eqnarray}
S_{BCS}&=&\sum_{ {\rm p},\sigma}(i\omega-\frac{p^2}{2 m}+\mu)
\fcreate{\sigma}{ \rm{p}}\fdestroy{\sigma}{ \rm{p}}\label{one}\\
&-&\frac{1}{\beta V}\hspace{-5mm}\sum_{ {\rm p}_1+ {\rm p}_2= {\rm p}_3+ {\rm p}_4}
\hspace{-5mm}U \fcreate{\uparrow}{ \rm{p}_1}\fcreate{\downarrow}
{ \rm{p}_2}\fdestroy{\downarrow}{ \rm{p}_3}\fdestroy{\uparrow}{ \rm{p}_4}\nonumber.
\end{eqnarray}
The remaining part of the action we will label the molecular action
\begin{eqnarray}
S_{M}&=&\sum_{ {\rm q},j}(iv-\frac{q^2}{4m}-\nu_j+2\mu)
\bcreate{ \rm{q}}\bdestroy{ \rm{q}}\label{two}\\
&&\hspace{-5mm}-\frac{1}{\sqrt{\beta V}}\hspace{-2mm}
\sum_{ {\rm q}= {\rm p}_1+{\rm p}_2, j}\hspace{-2mm}g_j\left(\bcreate{\rm{q}}
\fdestroy{\downarrow}{ \rm{p}_1}\fdestroy{\uparrow}{ \rm{p}_2}+
\fcreate{\uparrow}{ \rm{p}_2}\fcreate{\downarrow}{ \rm{p}_1}
\bcreate{ \rm{q}}\right).\nonumber
\end{eqnarray}
In deriving Eqs.~(\ref{one}) and (\ref{two}) we have inserted contact potentials for the couplings $U(\boldvec{x}-\boldvec{x}')\rightarrow U\delta(\boldvec{x}-\boldvec{x}')$ and 
$g_j(\boldvec{x}-\boldvec{x}')\rightarrow g_j\delta(\boldvec{x}-\boldvec{x}')$.  The full partition function
for our resonant system, under the model Hamiltonian of
Eq.~(\ref{ham}) can now be written as
\begin{equation}
Z=\int\left(\prod_\sigma {\rm D}a_\sigma^*{\rm D} a_\sigma\right)\left(\prod_j {\rm D}b_j^* 
{\rm D}b_j^{\phantom *}\right)\  {\rm e}^{S_{BCS}+S_M}\label{part},
\end{equation}
with the functional integral, ${\rm D}c\equiv \prod_i{\rm}dc^i$, ranging over all Fermi and Bose fields.

\section{Saddle-point approximation} 

From the form of the action in Eq.~(\ref{part}), it should be apparent
that all of the resonant contributions are contained within the
molecular action. In practice this gives rise to the integral of a
displaced Gaussian that can be easily evaluated. After integrating out
the molecular degrees of freedom, we are left with the partition
function:
\begin{equation}
Z=\left(\prod_j Z_{B_j}(q_j^2/4m+\nu_j-2\mu)\right)
\int{\rm D} a_\sigma^*{\rm D} a_\sigma{\rm e}^{S_{BCS'}}.
\label{partmod}
\end{equation}
Here $Z_{B_j}(q_j^2/4m+\nu_j-2\mu)$ is a Bose partition function
describing the formation of bound molecules and $S_{BCS'}$ is the BCS
action with a potential that is now dependent on both thermal
frequencies and momentum.  The interaction potential in the BCS
action is, therefore,
modified in the presence of a Feshbach resonance in the following way:
\begin{equation}
U\rightarrow U-\sum_j\frac{g_j^2}{q_j^2/4m+\nu_j-2\mu-iv}.
\end{equation}
With the above partition function, Eq.~(\ref{partmod}), we may go on to  
calculate all thermodynamic properties of interest. Here, we are primarily interested in calculating the
critical temperature of the superfluid phase transition. This can be
done by solving for the gap and number equation, and then
self-consistently solving these two equations for both the chemical
potential and the critical temperature. The procedure is straightforward since the full resonant calculation has been
reduced to the usual BCS calculation, only with a more complicated
potential. Following Popov's derivation \cite{popov}, introducing the
complex auxiliary Bose field $c({\rm q})$ and expanding about the
neighborhood of its zero value (which is equivalent to saying that we
expand about the zero of the gap near $T_c$), we derive the gap
equation at the critical point
\begin{equation}
1=\left(-U+\sum_j\frac{g_j^2}{\nu_j-2\mu}\right)\sum_k
\frac{\tanh(\beta(k^2/2m-\mu)/2)}{2(k^2/2m-\mu)}\label{gap}.
\end{equation}
The second self-consistent equation, the number equation, is found in
the saddle-point approximation by expanding the action to lowest order, i.e. 
$c({\rm q})=c^*({\rm q})=0$, and using the thermodynamic identity
$N=-\partial \ln Z/ \partial \mu$ giving
\begin{eqnarray}
N &=&2\sum_{j,k} \frac{1}{e^{\beta(k^2/4m+\nu_j-2\mu)}-1}
+2\sum_k \frac{1}{e^{\beta(k^2/2m-\mu)}+1}\label{num1}.\nonumber\\
\end{eqnarray}
Thus, our number equation counts all free fermions, $N_f$, plus an additional boson population $N_b$.  Equations~(\ref{gap}) and (\ref{num1}) provide for us a set
of equations for determining $T_c$ and $\mu$ at the critical point.
In the usual BCS theory, this level of approximation proves reasonable
for calculating $T_c$ in the BCS limit (small negative scattering
length) but diverges as the scattering length grows, and is wholly
inapplicable for positive scattering lengths.  The reason for this is
that the primary mechanism for the phase transition within the weak
coupling BCS limit is the formation and disassociation of Cooper
pairs.  As the coupling increases the particles tend to pair up at
higher and higher temperatures which means that the critical transition is no
longer signaled by the formation of Cooper pairs, but rather by a
coherence across the sample caused by condensation of preformed Cooper
pairs.  Since we are interested in describing the resonant system at
all detunings, the equations that we have derived so far are
insufficient because they do not account for this process. We should,
therefore, next focus on how to more accurately incorporate atom
pairing into our model.

\section{Beyond the saddle-point approximation}

To account for fluctuations in the fermion field, we follow
the method of Nozi\`eres and Schmitt-Rink \cite{nozieres} in its
functional form as put forth by Randeria {\it et\ al.}~\cite{randeria}.
This procedure will introduce a next order correction to the saddle
point calculation of the previous section.  By expanding the action to
second order in $c({\rm q})$, and then calculating the number equation
in the same way as was done when deriving Eq.~(\ref{num1}), we
introduce an additional population into the equation. The action
becomes
\begin{equation}
S(c({\rm q}),c^*({\rm q}))_{BCS'}\approx S_{BCS'}(0,0)+\sum_{{\rm q}}|c({\rm q})|^2\chi({\rm q}),
\end{equation}
where we have defined the auxiliary function $\chi({\rm q})$ as
\begin{eqnarray}
\hspace{-10mm}\chi({\rm q})=\left(U-\sum_j\frac{g_j^2}
{\frac{q^2}{4m}+\nu_j-2\mu-i\omega}\right)\sum_k
\frac{1-f(\epsilon_{\frac{q}{2}+k})-f(\epsilon_{\frac{q}{2}-k})}
{\epsilon_{\frac{q}{2}+k}+\epsilon_{\frac{q}{2}-k}-i\omega}.
\nonumber\hspace{-8mm}\\
\end{eqnarray}
Here $f(\epsilon_k)$ is the Fermi distribution function and \mbox{$\epsilon_k=k^2/2m-\mu$}.  The resulting modified number equation is
\begin{eqnarray}
N&=&2\sum_{j,k} \frac{1}{e^{\beta(k^2/4m+\nu_j-2\mu)}-1}
+2\sum_k \frac{1}{e^{\beta(k^2/2m-\mu)}+1}\nonumber\\
&-&\frac{1}{\beta}\sum_{\smallvec{q},\omega}\frac{\partial}
{\partial \mu}\log[1-\chi(\boldvec{q},i\omega)].\label{modnum}
\end{eqnarray}
This inclusion of the first order fluctuations
introduces a population of atom pairs, $N_p$,
that behave like bosons.  We are now able to solve for the fluctuation corrected critical temperature from a self-consistent solution of Eqs.~(\ref{gap}) and (\ref{modnum}).

Due to the contact form of the couplings
that we have chosen, however, we are immediately plagued with problems of divergences in our equations.  This can be
remedied by a proper renormalization, which means replacing the `bare'
couplings and detunings by the correct renormalized forms depending on the actual physical parameters as well as a momentum cut-off $K_{cut}$.  This same procedure is needed to
renormalize the usual, nonresonant BCS theory \cite{abrikosov} and is done by
relating the bare potential to the two-particle scattering matrix, $T$, through the
Lippman-Schwinger equation.  This leads to a renormalization of the coupling as $U=\Gamma \bar{U}$, where the bar notation represents a bare parameter.  We make the definition $\alpha=m K_{cut}/(2\pi^2)$ and introduce the dimensionless factor $\Gamma=(1-\alpha \bar{U})^{-1}$. This renormalization incorporates the true microscopic
physics into the problem, removing the unphysical divergence.

For the resonant case, a more sophisticated approach is required,
which we have extensively studied previously~\cite{servaas}. In the
case of a single resonance (i.e.,\ the $j$ sum has only a single
term) the renormalization is performed by
the following relations: $U=\Gamma \bar{U}$, $g =\Gamma \bar{g} $,
and $\nu =\bar{\nu} +\alpha g  \bar{g} $. The chemical potential
of each atomic state is modified to $\mu=\bar{\mu}-<T(k) n_k>$ to include
the proper mean-field shifts induced by all two-body scattering
processes, where $n_k$ is the Fermi distribution, $T(k)$ is the two-body
scattering matrix for the resonant system \cite{servaas}, and $<>$ denotes an averaging. This shift, however, is sufficiently small to neglect; inclusion has demonstrated corrections of the order of 1\% or less. By
replacing the bare values in Eqs.~(\ref{gap}) and (\ref{modnum})
with the renormalized values, all of the results to be discussed have
been shown to be independent of the introduced momentum cutoff
$K_{cut}$.

\begin{figure}[t]
  \includegraphics[scale=.37]{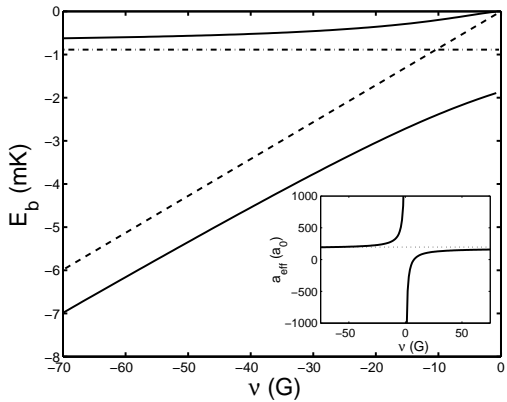}
\caption{ Binding energies for $^{40}{\rm K}$ resonance at positive background scattering length ${\rm a}_{\rm bg}=176 a_0$. A single resonance with a positive background scattering  length produces an effective scattering length ${\rm a}_{\rm eff}={\rm a}_{\rm bg}(1-\kappa /\nu )$, as seen in the figure inset where we plot the effective scattering length vs. detuning (the dotted line is at $176 a_0$). A positive $a_{\rm bg}$, which is larger than the range of the potential, implies that another bound state is not far below threshold (dash-dotted line). In combination with the Feshbach state (dashed detuning line) this results in an avoided crossing and the molecular state of interest asymptotes quickly to the dash-dotted line.}
\label{scattfig1}
\end{figure}

\begin{figure}[t]
  \includegraphics[scale=.37]{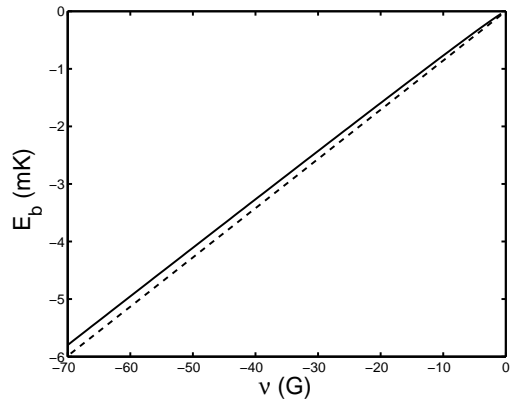} 
\caption{
 Same as Fig.~\ref{scattfig1}, but now for an artifical situation with $a_{{\rm bg}}<0$. A resonance system with a negative background scattering length has only one bound state relatively close to threshold, which is shifted positive of the detuning.  The next bound state in the potential is too far away to be of any significant influence.  The line styles are the same as for Fig.~\ref{scattfig1}.}
\label{scattfig2}
\end{figure}

The renormalization of the resonance theory forces us to take a closer look at the bound state physics of the system. In Fig.~(\ref{scattfig1}) we show the bound state energies for a single resonance system with a positive background scattering length. The figure results from a coupled square well calculation of the bound state energies \cite{servaas}, and shows the avoided crossing of two molecular states.   The upper state behaves to a fairly good approximation as $E_b=(ma_{\rm eff}^2)^{-1}$, which is the molecular binding energy regularly associated with a contact interaction \cite{newton}.  The lower state, however, is offset from the detuning by an energy $\sim \kappa $ and goes linear with the detuning. We find a similar behavior as in the lower state in the first term of Eq.~(\ref{modnum}). Taking the cutoff to infinity, which is justified since this term does not diverge, the
renormalized detuning approaches $\nu \rightarrow
\bar{\nu} -\bar{g} ^2/\bar{U}$. This produces a constant shift of  $\bar{g} ^2/\bar{U}=\kappa $ between the detuning and the molecular binding energy. Keeping this term in the number equation would incorrectly cause a transfer of the entire population into the wrong molecular state.  In order to avoid this unwanted behavior we set this term to zero, i.e. $N_b=0$.  In the case of a negative background scattering length, we would not have encountered this problem, and only one molecular state would have appeared (see Fig.~\ref{scattfig2}). We will show in the next section that the pairing term fully accounts for the correct population of molecules in this system with $a_{\rm bg}>0$.

Before we present the full crossover solution for the case of $^{40} \rm{K}$, let us
look at the analytical solutions to Eqs.~(\ref{gap}) and (\ref{modnum}) in the strong (BEC) and weak (BCS)
coupling regimes.  We will first turn our attention to the weak
coupling (BCS) regime. In this limit we would expect only free
fermions to contribute to the population, so from equation (\ref{modnum})
we find that the chemical potential is at the Fermi surface
($\mu=E_{\rm F}$).  With this information, we solve the gap equation
for the critical temperature.  The result is the usual exponential
dependence on the effective scattering length
\begin{equation}
T_c/T_F\approx\frac{8}{\pi}e^{\gamma-2}\exp(\frac{-\pi}{2 k_{\rm f} 
|{\rm a}_{\rm eff}|}),
\end{equation}
where $\gamma\sim 0.5772$ is the Euler-Mascheroni constant, $k_{\rm f}$
is the Fermi wave number, and ${\rm a}_{\rm eff}<0$ is the effective scattering
length produced by the Feshbach resonance
${\rm a}_{\rm eff}=a_{\rm bg}(1-\kappa /\nu )$.

The other limit we may consider is the strong coupling (BEC)
limit.  When the argument of the $\tanh$ function in the gap equation (\ref{gap}) 
becomes sufficiently negative, it is a good approximation to use its
asymptotic value of unity. What this means physically is that the
fermion statistics are unimportant in determining the value of the
gap. This allows us to solve the gap equation for the chemical
potential as a function of detuning.  In the limit of large negative
detuning we find that $\mu\rightarrow -E_b/2$, where $E_b\approx 1/ma_{\rm eff}^2$. Within this limit
the entire population has been converted to molecules and we can solve
the number equation to get the BEC condensation temperature of
$T_c/T_F\sim0.218$. 

\section{Numerical results} 

To study the transition between the BEC and BCS regimes, we numerically solve Eqs.~(\ref{gap}) and
(\ref{modnum}) for $^{40} \rm{K}$.  The single resonance curve is
produced using a background scattering length of $176 a_0$ and
$\kappa =7.68 {\rm G}$ at
a density of $10^{14}{\rm cm}^{-3}$.

\begin{figure}
  \includegraphics[scale=.37]{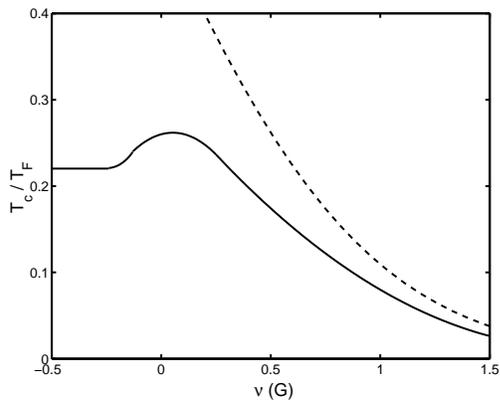}
\caption{
  Critical temperature $T/T_F$ as a function of detuning $\nu$ in
  gauss.  The dashed line corresponds to the usual BCS solution, which limits to the full crossover theory at large positive detuning. At negative detuning, $T_c$ drops to the BEC condensation temperature of $T_c/T_F\sim0.218$.}
\label{tcfig}
\end{figure}

\begin{figure}
  \includegraphics[scale=.37]{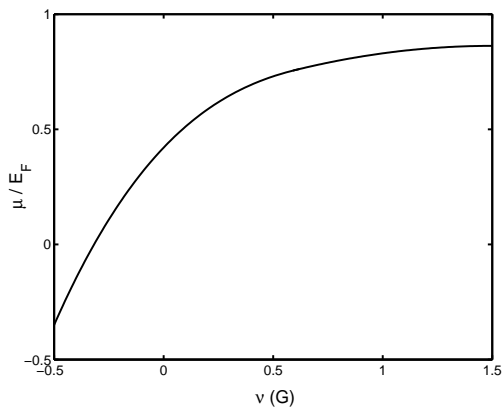} 
\caption{
  Chemical potential as a function of detuning $\nu$ in gauss. For large negative detuning $2\mu$ approaches the bound state energy of the molecular state. At increasing positive detuning, the chemical potential slowly approaches the Fermi energy.}
\label{chemfig}
\end{figure}

\begin{figure}
  \includegraphics[scale=.37]{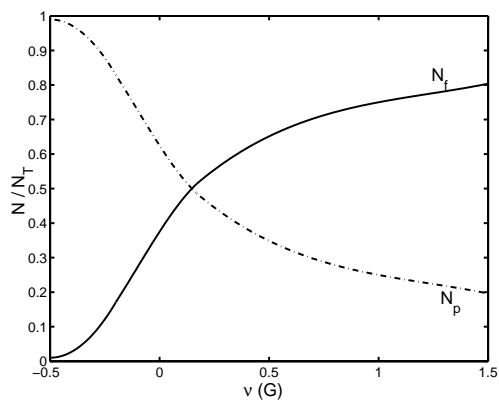} 
\caption{The fraction of the total population as a function of detuning $\nu$ in gauss. The dot-dashed line corresponds to the pairing fraction $N_p$ and the solid to the free fermion fraction $N_f$.}
\label{numfig}
\end{figure}

Figure~\ref{tcfig} shows the critical temperature as a function of
magnetic field detuning.  The crossover calculation clearly merges with the BEC result for large-positive detunings and
smoothly connects between positive and negative detunings, limiting to the Bose condensation temperature of $T_c/T_F\sim0.218$ for large negative detuning. This
approach gives a maximum  near zero detuning ($T_c/T_F\sim0.26$) , but the maximum critical
temperature we find is less than the predictions of the HFB approach
in our earlier papers ($T_c/T_F\sim0.5$) \cite{chiofalo}. We believe this is due to the
inclusion of fluctuations in the beyond saddle point approximation
which act to reduce the gap at zero temperature and therefore the
critical temperature for the formation of a superfluid.

Figure~\ref{chemfig} shows the chemical potential as a function of
detuning, beginning at the Fermi energy for positive detuning and
approaching half the bound state energy at large negative detuning
\mbox{$\mu\rightarrow -E_b/2$}.  Figure~(\ref{numfig}) shows the change in population as a function of
detuning.  For large positive detuning, the system is composed solely
of free fermions.  As the detuning is decreased (i.e. from positive to negative) the contribution of the fermions begins to decrease until all the population is transferred into the atom pairs at $\nu\sim-0.5 G$.  The chemical potential is then equal to $-E_b/2$ and we may identify the atom pairs from that point as the molecules.  The
superfluid behavior then comes from the condensation of these molecules, which are no longer disassociating into free fermions.  

\section{Conclusion} 
We have presented a crossover model to describe the
behavior of a gas of fermionic atoms for all detunings
from a Feshbach resonance.  The model is able to smoothly connect between the
BCS and BEC regimes and accounts for the microscopic two-body
physics throughout. We find a smooth behavior of the transition temperature in the entire
crossover regime, with a maximum near zero detuning of $T_c/T_f\approx
0.26$, and agreeing with the appropriate BEC and BCS behaviors in the
respective large detuning limits. The maximum is below the value
predicted by the Hartree-Fock-Bogoliubov theory for the uniform gas
derived in earlier work. This result is a direct indication of the
important role of preformed atom pairs which are neglected above $T_c$ in
the Hartree-Fock-Bogoliubov theory. Such atomic pairing is represented as
fluctuations in the fermion pairing field and modify the elementary
excitation spectrum even in the normal phase. We emphasize, however,
that in this paper we have only accounted for the pairing physics by
including the second order fluctuations in order to be able to account
for the correct molecular binding energy as derived from the two-body
scattering physics. While the inclusion of this order of fluctuations is
the main ingredient necessary to encapsulate essential aspects of the
behavior of the system in the crossover regime, it would be
interesting to extend the approach to consider the effect of higher
order interactions which have not been accounted for.  For example, we
have only included interactions between free fermion atoms, neglecting
all other contributions such as the interactions between pairs. This
is most clearly seen in the BEC limit, where our solution adopts the
thermodynamic behavior of the ideal Bose gas, rather than the dilute
interacting Bose gas. A more sophisticated treatment could extend our
approach to consider all these factors. Nevertheless, the results we
have presented here illustrate the realistic potential in this
realizable system for increasing the superfluid transition temperature, with the aid of the Feshbach resonance, to a significant fraction of
the Fermi temperature. This is an important and timely aspect from a
practical perspective, because the maximum value we predict is in the
region of the temperature range which is currently experimentally
accessible.

\begin{acknowledgments}
Support is acknowledged for J. M. and S. K. from the U. S. Department of Energy, Office of Basic Energy Sciences via the Chemical Sciences, Geosciences and Biosciences Division, and for M. H. from the National Science Foundation.
\end{acknowledgments}

\end{document}